\documentclass[11pt,a4paper]{article}
\pdfoutput=1
\usepackage{jcappub}

\usepackage{epstopdf}
\usepackage{natbib}
\usepackage[percent]{overpic}

\def\({\left(}
\def\){\right)}

\def\integ0a{\int_0^a}
\def\[{\left[}
\def\]{\right]}

\newcommand{\ang}{\hat\nabla}
\newcommand{\up}[1]{{\rm #1}}
\newcommand{\bdv}[1]{{\bf #1}}

\newcommand{\beeq}{\begin{equation}} 
\newcommand{\eneq}{\end{equation}}
\newcommand{\bear}{\begin{eqnarray}}
\newcommand{\enar}{\end{eqnarray}}

\newcommand{\vL}{{\bf l}}
\newcommand{\Vang}{\bdv{\hat n}}

\newcommand{\lenX}{\tilde X}

\newcommand{\CT}{C^T}
\newcommand{\CE}{C^E}

\newcommand{\CC}{C^C}
\newcommand{\lenCX}{\tilde C^X}

\newcommand{\Xobs}{\tilde X^{\up{obs}}}
\newcommand{\CXobs}{\tilde C^{X,\up{obs}}}
\newcommand{\Yobs}{\tilde Y^{\up{obs}}}
\newcommand{\CYobs}{\tilde C^{Y,\up{obs}}}
\newcommand{\GXY}{\bdv{G}_{XY}}
\newcommand{\WY}{W_Y}
\newcommand{\khat}{\hat\kappa}
\newcommand{\Ckap}{C^{\kappa\kappa}}
\newcommand{\Nkap}{N^{\kappa\kappa,XY}}

\newcommand{\spix}{\sigma_{\up{pix}}}
\newcommand{\opix}{\Omega_{\up{pix}}}

\newcommand{\sbeam}{\sigma_b}
\newcommand{\tfwhm}{\theta_{\up{FWHM}}}

\newcommand{\bn}{\hat{\bf n}}
\newcommand{\bl}{{\bf l}}
\newcommand{\bll}{{\bf L}}

\newcommand{\intlp}[1]{\int {d^2 l_{#1}' \over (2\pi)^2}}

\newcommand{\WMAP}{\textsl{WMAP}}

\newcommand{\eqn}[1]{Eq.~(\ref{#1})}

\def\be{\begin{equation}}
\def\ee{\end{equation}}
\def\bea{\begin{eqnarray}}
\def\eea{\end{eqnarray}}

\def\simlt{\lower.5ex\hbox{$\; \buildrel < \over \sim \;$}}
\def\simgt{\lower.5ex\hbox{$\; \buildrel > \over \sim \;$}}
\def\simgtalt{\lower.5ex\hbox{$\buildrel > \over \sim \;$}}

\def\aa{{\sl Astron.\ \&\ Astrophys.\ }}

\def\apj{{\sl Astrophys.\ J.\ }}

\def\apjs{{\sl Astrophys.\ J.\ Supp.\ }}

\def\jcap{{\sl J.\ Cosm.\ Astroparticle\ Phys.\ }}

\def\mnras{{\sl MNRAS\ }}

\def\physrep{Phys. Rept.\ }

\def\prd{{\sl Phys.\ Rev.\ D\ }}

\title{N-body lensed CMB maps: lensing extraction and characterization}

\author[a,b]{Claudia Antolini,}
\author[c,d]{Yabebal Fantaye,}
\author[a,b]{Matteo Martinelli,}
\author[e,f]{Carmelita Carbone,}
\author[a,b]{Carlo Baccigalupi}
\affiliation[a]{SISSA, Via Bonomea 265,  I-34136 Trieste, Italy}
\affiliation[b]{INFN, Sezione di Trieste, Via Valerio 2, I-34127 Trieste, Italy}
\affiliation[c]{Institute of Theoretical Astrophysics, University of Oslo, P.O. Box 1029 Blindern, N-0315 Oslo, Norway}
\affiliation[d]{Department of Mathematics, University of Rome Tor Vergata, Via della Ricerca Scientifica 1, I-00133 Rome, Italy}
\affiliation[e]{INAF-Osservatorio Astronomico di Brera, Via Bianchi 46, I-23807 Merate (LC), Italy}
\affiliation[f]{INFN, Sezione di Bologna, Viale Berti Pichat 6/2, I-40127 Bologna, Italy}
\emailAdd{claudia.antolini@sissa.it}
\emailAdd{y.t.fantaye@astro.uio.no}
\emailAdd{mmartin@sissa.it}
\emailAdd{carmelita.carbone@brera.inaf.it}
\emailAdd{bacci@sissa.it}
\abstract{We reconstruct shear maps and angular power spectra from simulated weakly lensed total intensity ($TT$) and polarised ($EB$) maps of the Cosmic Microwave Background (CMB) 
anisotropies, obtained using Born approximated ray-tracing through the N-body simulated Cold Dark Matter (CDM) structures in the Millennium Simulations (MS). We compare the recovered 
signal with the $\Lambda$CDM prediction, on the whole interval of angular scales which is allowed by the finite box size, extending from the degree scale to the arcminute, 
by applying a quadratic estimator in the flat sky limit; we consider {\sc PRISM}-like instrumental specification for future generation CMB satellites, corresponding to arcminute 
angular resolution of $3.2'$ and sensitivity of $2.43$ $\mu$K-arcmin. The noise contribution in the simulations closely follows the estimator prediction, becoming dominated
by limits in the angular resolution for the $EB$ signal, at $\ell\simeq 1500$. The  recovered signal shows no visible departure from predictions of the weak lensing power 
within uncertainties, when considering $TT$ and $EB$ data singularly. In particular, the reconstruction precision reaches the level of a few percent in bins with 
$\Delta\ell\simeq 100$ in the angular multiple interval $1000\lesssim\ell\lesssim 2000$ for $T$, and about $10\%$ for $EB$. Within the adopted specifications, polarisation data 
do represent a significant contribution to the lensing shear, which appear to faithfully trace the underlying N-body structure down to the smallest angular scales achievable 
with the present setup, validating at the same time the latter with respect to semi-analytical predictions from $\Lambda$CDM cosmology at the level of CMB lensing statistics. 
This work demonstrates the feasibility of CMB lensing studies based on large scale simulations of cosmological structure formation in the context of the current and future high 
resolution and sensitivity CMB experiment.}
\arxivnumber{1311.7112}

\begin{document}

\maketitle

\newpage
\section{Introduction}
\label{sec:i}

The anisotropies in the Cosmic Microwave Background (CMB) are one of the most important observables of modern cosmology. Their study substantially contributed to the establishment of a reference 
cosmological model consistent with a flat Friedmann Robertson Walker metric and made of three main components, baryons and leptons representing about $4\%$ of the total 
energy density, Cold Dark Matter (CDM, about $26\%$) which constitutes the large part of the gravitational potential around cosmological structures, and about $70\%$ of a Dark 
Energy (DE) component, similar or coincident with a Cosmological Constant ($\Lambda$, CC), responsible for a late time phase of accelerated expansion. The early Universe phenomenology 
is consistent with the inflationary picture, with a Gaussian and almost scale invariant initial perturbation spectrum dominated by scalar modes. Modern observations allow 
to measure these quantities to percent precision or better (see \cite{planck_mission_paper} and references therein).\\
More and more efforts are being undertaken for measuring second order effects, i.e. physical phenomena which occur after the last scattering surface,  the epoch at which 
CMB photons decouple from the rest of the system and the first order anisotropies are imprinted. In order to constrain the dark cosmological components, and the DE in particular, 
the observation and characterization of the weak lensing of the CMB induced by forming structures along the line of sight of photons at the epoch in which the DE overcomes the CDM 
component \citep{acquaviva_baccigalupi_2006} is gathering more and more attention. CMB lensing, in fact, allows us to  break the degeneracies present in the measurements of 
cosmological parameters through CMB observations only~\citep{Stompor:1999mnrs} as well as providing more constraining power on the same 
parameters  ~\citep{Perotto_FutureCMB_06,2012arXiv1205.0474B}. Moreover, as lensing is closely related to the underlying gravitational theory, it can be used to test the 
possibility that the late-time accelerated expansion is not given by a DE component, but rather by a modified theory of gravity~\cite{Calabrese:2009tt}. As CMB lensing is a 
second order effect in cosmological perturbations, sourced by forming structures onto primary anisotropies, it acts on the total intensity ($T$) anisotropies through a smearing of the 
acoustic peaks on sub-degree angular scales, as well as the transfer of power to the arcminute scale \citep{bartelmann_schneider_2001}; the same effect is induced on the scalar-type 
mode of CMB polarisation, the $E-$mode, while a fraction of power in the latter leaks to the curl component, the $B-$mode, causing a characteristic peak centered on the scales 
of a few arcminutes. Furthermore, since primordial gravitational waves contribute to the $B$ polarization modes on the degree scale, the latter effect in particular from 
gravitational lensing must be correctly taken into account when information about the ratio between tensor and scalar modes in cosmology ($r$) is to be 
obtained (see \cite{antolini_etal_2012} and references therein). \\
This CMB lensing signal, first detected correlating the data from the Wilkinson Microwave Anisotropy Probe (\WMAP) with matter tracers~\cite{2007PhRvD..76d3510S}, has been then 
measured through CMB observations alone by the ACT collaboration \citep{das2011}, the South Pole Telescope (SPT)~\citep{Engelen:2012spt} and with spectacular 
confidence in the first release of the {\sc Planck} data \citep{planck_xv_2013}; future CMB observations from the {\sc Planck} second data release and other sub-orbital experiments 
are expected to provide more and more precise measurements of this signal; recently, the cross-correlation of the $B-$modes as predicted by the cross-correlation between the 
SPT measurement of the $E-$mode polarisation with Herschel data and the actual polarisation $B$ measurement from SPT itself yielded a detection of the lensing $B-$modes to high 
confidence \citep{hanson_etal_2013}. 

In this scenario, our capability of understanding the lensing signal to extreme accuracy is most important and a necessary condition for that is to be able to model it appropriately 
and to the accuracy needed by modern cosmological observations. In the recent past, efforts were made in order to simulate lensed CMB maps in the context of modern N-body simulations, 
which, once validated, have the potential and crucial capability of enabling us to estimate the constraining power which we will have from CMB lensing in particular on the underlying cosmological model \citep[see][and references therein]{carbone_etal_2013} and most importantly in view of cross-correlating CMB lensing measurements with those of the Large Scale 
Structures (LSS) which are responsible for the lensing itself, culminating with the launch of the Euclid satellite in about one decade \citep{euclid_reference}. 
In this paper we progress on this line, by extracting and characterizing, for the first time, the lensing signal in simulated CMB temperature and polarisation maps using ray tracing through N-body 
simulations. We exploit a flat sky lensing extraction pipeline developed and exploited in \citep{Fantaye:2012ha} onto the CMB lensed maps which were constructed by performing 
ray tracing in the Born approximation using the Millennium Simulations (MS) in \citep{carbone_etal_2007,carbone_etal_2008}.\\

The paper is organized as follows. In Section~\ref{sec:ii} we introduce and discuss the details of the N-body simulations used to reconstruct the CMB maps, also specifying the methods 
used to produce these maps. In Section~\ref{sec:iii} we briefly review the theoretical background for lensing extraction methods and we detail the formalism and procedure 
followed. Section~\ref{sec:iv} contains the application of our extraction pipeline on the CMB maps reconstructed from the N-body simulations, assuming observational errors compatible 
with current and upcoming CMB surveys. We discuss our results in the concluding Section~\ref{sec:v}.

\section{From N-Body simulations to CMB maps}
\label{sec:ii}

Weak lensing of the CMB deflects photons coming from an original
direction ${\bf \hat{n}}'$ on the last scattering surface to a
direction ${\bf \hat{n}}$ on the observed sky, and the lensed CMB field
is given by $\tilde{X}({\bf\hat{n}})=X({\bf\hat{n}'})$ in terms of the
unlensed field $X=T,Q,U$. The vector ${\bf
  \hat{n}}'$ is obtained from ${\bf \hat{n}}$ by moving 
its end on the surface of a unit sphere by a distance
$|\nabla_{\bf\hat{n}}\phi({\bf \hat{n}})|$ along a geodesic in the
direction of $\nabla_{\bf\hat{n}}\phi({\bf \hat{n}})$, where
$\nabla_{\hat{\bf n}}$ is the angular derivative in the direction transverse
to the line-of-sight pointing along ${\hat{\bf n}}\equiv(\vartheta,\varphi)$
\citep{Hu:2000ee, Challinor02, Lewis05, Zaldarriaga:1998te}. Here the field
$\phi$ is the so-called ``lensing potential'', and
$|\nabla_{\bf\hat{n}}\phi({\bf \hat{n}})|$ is assumed to be constant between 
${\bf \hat{n}}$ and ${\bf \hat{n}}'$.
Therefore the lensed temperature and polarization fields are given by
\bear
\label{eq:lensing}
\tilde T(\bn)&=&T\left[\bn+\ang\phi(\bn)\right],\\
(\tilde{Q}+i\tilde{U})(\bn)&=&(Q\pm iU)\left[\bn+\ang\phi(\bn)\right].
\nonumber
\enar

In what follows we will consider only the \emph{small angle
  scattering} limit, \emph{i.e.} the case where the change in the
comoving separation of CMB light rays, owing to the deflection caused
by gravitational lensing from matter inhomogeneities, is small
compared to the comoving separation between the \emph{undeflected}
rays. In this case it is sufficient to calculate all the relevant
integrated quantities, \emph{i.e.} the lensing potential and
its angular gradient, the so-called \emph{deflection angle}, along the
undeflected rays. This small angle scattering limit corresponds to the
Born approximation.

Under this approximation, adopting conformal time and comoving coordinates \citep{maber}, the
integral for the projected lensing potential due to 
scalar perturbations in the absence of anisotropic stress reads
\bear
\label{lensingpotential}
\phi(\bn) = -2\int_0^{D_\star}dD~{\frac{D_\star-D}{ D D_\star}}~\psi~(D\bn,D)\,,
\enar
where $D$ and $D_\star$ are, respectively, the comoving angular
diameter distances to the lens and to the CMB last scattering surface,
and $\psi$ is the physical peculiar gravitational potential
generated by density perturbations
\citep{Hu:2000ee,Matthias_rew,Refregier,Lewis06}. Let us notice that
$\phi$ is connected to the convergence field $\kappa$ via
$\ang^2\phi=-2\kappa$. 

If the gravitational potential $\psi$ is Gaussian, the lensing
potential is also Gaussian. However, the lensed CMB is non-Gaussian,
as it is a second 
order cosmological effect produced by matter perturbations onto
CMB anisotropies, yielding a finite correlation between different
scales and thus non-Gaussianity.  This is expected to be most
important on small scales, due to the non-linearity already present in
the underlying properties of lenses.

In the present work we analyse the full sky $T$, $Q$, $U$ maps lensed by
the matter distribution of the MS and generated by
\cite{carbone_etal_2008} via a modification of the publicly available {\sc LensPix}
code\footnote{http://cosmologist.info/lenspix/} (LP), described in
\cite{Lewis05}. In its original version this
code lenses the primary CMB intensity and polarization fields using a
Gaussian realisation, in the spherical harmonic domain, of the lensing
potential power spectrum as extracted from the publicly available Code
for Anisotropies in the Microwave Background
(CAMB\footnote{http://camb.info/}).  The modification made by
the authors consists in forcing LP to deflect
the CMB photons using the fully non-linear and non-Gaussian lensing
potential realization obtained from MS exploiting the procedure
briefly summarized below, and presented in
\cite{carbone_etal_2008}; we refer the reader to this paper for further
details.

The MS is a high resolution $N$-body simulation for
a $\Lambda$CDM cosmology consistent with the \WMAP \ 1 year results
\citep{spergel2003}, carried out by the Virgo Consortium
\citep{Springel2005}. It uses about 10 billion collisionless particles
with mass $8.6\times 10^{8} h^{-1}{\rm M_\odot}$, in a cubic region
$500\,h^{-1}{\rm Mpc}$ on a side which evolves from redshift
$z_{*}=127$ to the present, with periodic boundary conditions.  The
map-making procedure developed in \cite{carbone_etal_2007} is based on
ray-tracing of the CMB photons in the 
Born approximation through the three-dimensional field of the MS
peculiar gravitational potential.  In order to produce mock lensing
potential maps that cover the past light-cone over the full sky, the
MS peculiar gravitational potential grids are stacked around the
observer located at $z=0$, and the total volume around the observer up
to $z_{*}$ is divided into spherical shells, each of thickness
$500h^{-1}{\rm Mpc}$: all the MS boxes falling into the same shell are
translated and rotated with the same random vectors generating a
homogeneous coordinate transformation throughout the shell, while
randomization changes from shell to shell. The peculiar gravitational
potential at each point along a ray in direction $\bf\hat{n}$ is
interpolated from the pre-computed MS potential grids which possess a spatial
resolution of about $195h^{-1}{\rm kpc}$.  

Being repeated on scales larger than the box size, the resulting weak
lensing distortion lacks large scale power, which manifests itself in
the lensing potential power spectrum as an evident loss of large scale
power with respect to semi-analytic expectations, most
noticeable at multipoles smaller than $l\simeq 400$. This has been
cured in \cite{carbone_etal_2007} by augmenting large scale power
(LS-adding) directly in the angular domain, a procedure which we
exploit here as well, since large 
scale modes in the lensing potential field are transferred to small
scales in the CMB field, causing, e.g., the increasing of the so-called
temperature damping tail with respect to the unlensed field. This mode coupling
effect, which produces the characteristic non-Gaussianity of the CMB
lensed field, is indeed exploited for the reconstruction of the
underlying matter deflecting field. Nonetheless, in this work, we are
mostly interested in studying the lensing reconstruction of the MS
matter field, which corresponds to scales $l>400$, and therefore,
while still using all sky CMB lensed maps as input, we will
exploit the flat sky lensing extraction pipeline for the recostructed lensing
potential output, as described in the next Section.

For the construction of the all sky lensed CMB input maps, in
\cite{carbone_etal_2008} the
LS-adding technique has been implemented 
directly into the LP code where the spherical harmonics domain has
been splitted into two multipole 
ranges: $0\leq l \leq 400$, where the MS fails in reproducing the
correct lensing potential power due to the limited box-size of the
simulation, and $l > 400$, where instead the power spectrum is
reproduced correctly.  On the latter
interval of multipoles, the corresponding ensemble
$\phi_{lm}^{\rm MS}$ of lensing potential spherical harmonic
coefficients produced by the MS lens distribution has been extracted.  
The LP code has been modified to read and use these MS harmonic
coefficients on the corresponding range of multipoles.  On the
interval $0\leq l \leq 400$, instead, LP generates its own
ensemble of spherical harmonic coefficients $\phi_{lm}^{\rm LP}$,
which are a realisation of a Gaussian random field characterised by
the CAMB semi-analytic non-linear lensing potential power spectrum
inserted as input in the LP parameter file. 

Since on multipoles $0\leq l \leq 400$ the effects of non-Gaussianity from the
non-linear scales are negligible and the $\phi_{lm}$ are independent,
every time that we run the MS-modified-LP, we generate a joined
ensemble of $\tilde{\phi}_{lm}$, where
$\tilde{\phi}_{lm}=\phi_{lm}^{\rm LP}$ for $0\leq l \leq 400$ and
$\tilde{\phi}_{lm}=\phi_{lm}^{\rm MS}$ for $l > 400$. This technique
reproduces correctly the 
non-linear and non-Gaussian effects of the MS non-linear dark matter
distribution on multipoles $l > 400$, including at the same time the
contribution from the large scales at $l \le 400$, where the lensing
potential follows mostly the linear trend.

To generate the lensed $T$, $Q$, $U$ fields from the MS-modified-LP
code, we adopt the method described in \cite{Lewis05}, using a high
value of the multipole $l_{\rm max}$ to maximize the accuracy.  This
allows running the 
simulation several times without excessive consumption of CPU time and
memory.  We work under the  assumption that tensor modes are
absent in the early Universe, so that the produced $B-$mode polarization
is due only to the power transfer from the primary scalar $E-$modes into
the lensing induced $B-$modes.  We choose $l_{\rm max}=6143$ and a
HEALPix\footnote{http://healpix.jpl.nasa.gov/} pixelisation parameter
$N_{\rm side}=2048$, which corresponds to an angular resolution of
$\sim 1.72^\prime$ \citep{Healpix}, with $12 N_{\rm side}^2$ pixels in
total.

\section{CMB Lensing Extraction}
\label{sec:iii}

As mentioned in the previous Section, we will work in the so-called ``flat-sky
approximation'' for the reconstruction of the lensing potential. In
this limit the lensing potential can be 
written as \citep{Zaldarriaga:1998te}:
\begin{eqnarray}
\phi(\bn) =\int { \frac{d^2 L }{(2\pi)^2}}\phi(\bll) e^{i \bll \cdot \bn}\, 
\end{eqnarray}
where the polar and azimuthal angles have been replaced by the displacement $\bl$. The corrections due to lensing in the Fourier moments of temperature and 
polarization fields can be expressed, at the linear order in $\phi$,
as~\citep{Hu:2000ee} 
\begin{eqnarray}
\delta \tilde T(\bl) &=& \intlp{} T(\bl') W(\bl',\bll),\\
\label{eqn:lensedl}
\delta \tilde E(\bl)      &=& \intlp{} 
\Big[ E(\bl') \cos 2\varphi_{\bl'\bl}
     - B(\bl') \sin 2\varphi_{\bl'\bl} \Big]
W(\bl',\bll),\nonumber\\
\delta \tilde B(\bl)      &=& \intlp{} 
\Big[ B(\bl') \cos 2\varphi_{\bl'\bl}
     + E(\bl') \sin 2\varphi_{\bl'\bl}\Big]
W(\bl',\bll),\nonumber
\end{eqnarray}	
where the azimuthal angle difference is $\varphi_{\bl'\bl} \equiv \varphi_{\bl'} - \varphi_{\bl}$,
$\bll = \bl - \bl'$, and
\begin{eqnarray}
W(\bl,\bll) = -[{\bl \cdot \bll}]\phi(\bll).
\end{eqnarray}
These equations show that lensing couples the gradient of the primordial CMB modes $\bl'$ to that of the observed modes $\bl$. Furthermore, even if primordial $B-$modes are 
vanishing, $ B(\bl')=0$, lensing generates $B-$mode anisotropies in the observed map given the leakage from the $E$ and $T-$modes.\\

We will consider noise in the CMB maps assumed homogeneous and white, characterized by a Gaussian beam. The power spectrum of the detector noise
is~\citep{Knox_95}
\begin{eqnarray}
C_{l}^{N,X}=\spix^2~\opix,
\label{eq:nois}
\end{eqnarray}
where $\spix$ is the r.m.s. noise per pixel and $\opix$ is the solid angle subtended by each pixel. The observed CMB temperature and polarization fields, $X\in[T,E,B]$, and their power spectra, $\tilde C_{\ell}^X$, are \bear
\label{eq:tobs}
\Xobs_\vL&=&\lenX_\vL~e^{-{\frac12}l^2\sbeam^2}+N^X_\vL,\\
\CXobs_l&=&\lenCX_le^{-l^2\sbeam^2}+C^{N,X}_l, \nonumber
\enar
where $N^X_\vL$ is the Fourier mode of the detector noise, and $\sbeam$ is related to the Full Width at Half Maximum (FWHM) of the telescope beam, $\theta_{\rm FWHM}$, as $\tfwhm=\sbeam\sqrt{8\ln2}$.\\

\begin{figure}[t]
\begin{center}
\includegraphics[width=10cm]{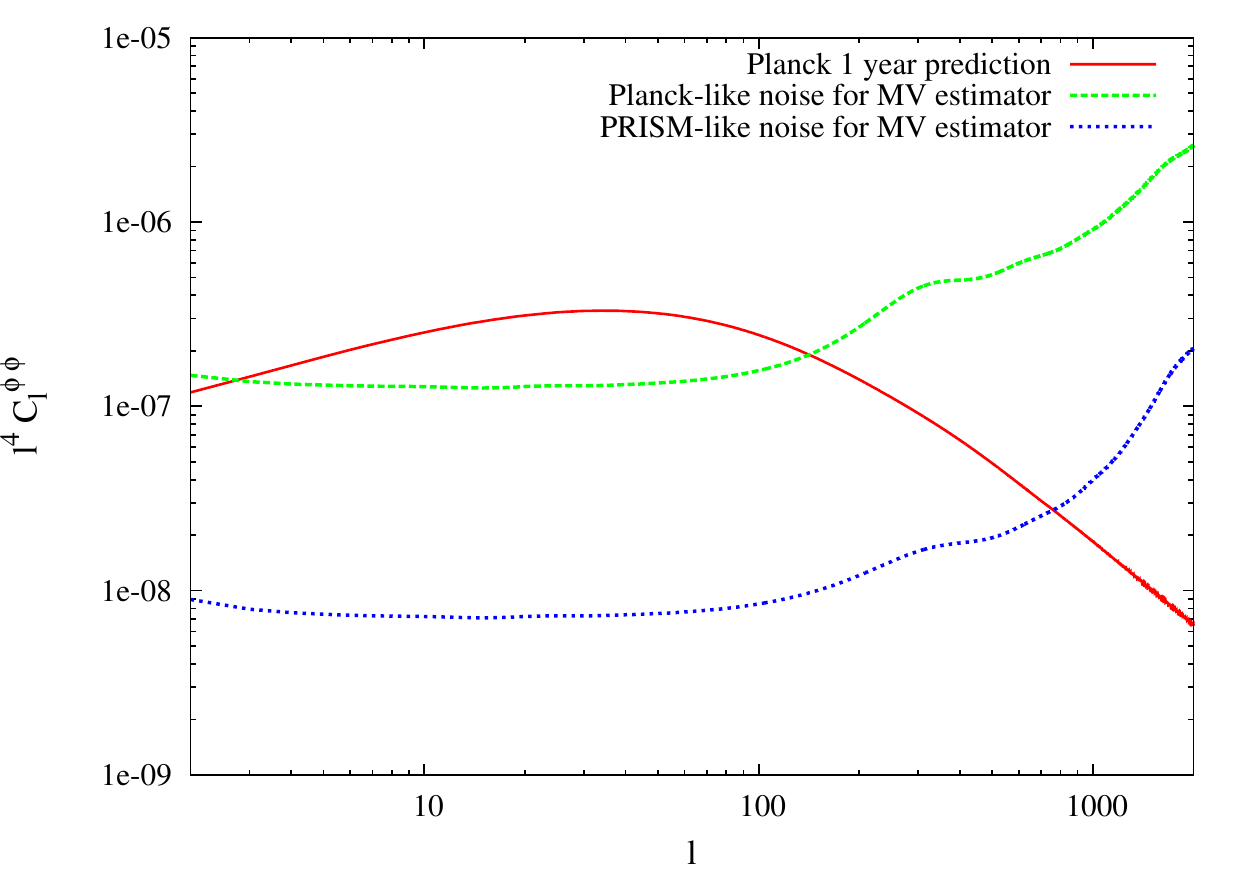}
\caption{ Noise spectrum for different experimental specifications. For graphical purposes we show the convergence power spectrum which is connected to the potential through $C_\ell^{\kappa \kappa}=l^4C_\ell^{\phi \phi}$. In this figure, the prediction for the convergence spectrum using as input cosmology the estimated cosmological parameters coming from
{\sc Planck} 1 year observations (red solid line), the noise for the minimum variance quadratic estimator for a {\sc Planck}-like experiment (green dashed line) and the the noise for the same estimator for a {\sc PRISM}-like experiment (blue dashed line) are plotted. }
\label{figure_noise}
\end{center}
\end{figure}

We exploit the quadratic estimator formalism~\citep{Hu:2001fa,Hu:2001tn,HuOkamoto:2002}, built  in the context of the convergence estimators~\citep{Hu:2007njp, 2010PhRvD..81l3006Y}, 
in order to extract the lensing information from the simulated CMB maps used in the analysis.\\
The estimator is uniquely determined by requiring each component to be unbiased over an ensemble average of the CMB temperature and polarization fields $X$~and~$Y$ ($\langle\khat^{XY}(\Vang)\rangle=\kappa(\Vang)$) and the variance of the estimator to be minimal, 
\begin{eqnarray}
\langle\khat^{XY}_\vL\khat^{*XY}_{\vL'}\rangle=(2\pi)^2~\delta^D(\bdv{l-l'})
(\Ckap_l+\Nkap_l)\ , 
\label{eq:vari}
\end{eqnarray}
 where the $\Nkap_l$ term represents the noise contribution which is also predicted by the estimator, as we see below.
In real space the convergence estimators are constructed on the basis of appropriate filters of the observed fields, weighted in the harmonic domain by their power spectra, 
which are given by
\cite{2010PhRvD..81l3006Y}
\bear
\label{eq:filter}
&&\GXY(\Vang)=\int\!\!\!{\frac{d^2\vL}{(2\pi)^2}}~i\vL\Xobs_\vL
{\frac{C_l^{XY}}{\CXobs_l}}
\bigg\{\begin{array}{c}e^{2i\varphi_{\vL}}\\e^{2i\varphi_{\vL}}\end{array}\bigg\}
e^{-{\frac12}l^2\sbeam^2+i\vL\cdot\Vang}\\
&&\WY(\Vang)=\int\!\!\!{\frac{d^2\vL}{(2\pi)^2}}~{\frac{\Yobs_\vL}{\CYobs_l}}
\bigg\{\begin{array}{c}e^{2i\varphi_{\vL}}\\ie^{2i\varphi_{\vL}}\end{array}\bigg\}
e^{-{\frac12}l^2\sbeam^2+i\vL\cdot\Vang}\label{eq:filter2}
\enar
where $\varphi_\vL$ is the azimuthal angle of the wave vector~$\vL$; the two phase factors in braces are applied when $Y=E,B$ respectively, and are unity when $Y=T$.  Also, $C_l^{XY}=C_l^{XE}$ for $Y=B$. In the construction of these fields the map beam deconvolution is incorporated, hence the beam factors $e^{-{\frac12}l^2\sbeam^2}$ appearing on both fields.\\ 
Given the two filtered fields in \eqn{eq:filter} and \eqn{eq:filter2}, the convergence estimators are then given by \begin{eqnarray}
\khat^{XY}_\vL=-\frac{A^{XY}_l}{2}~i\vL\cdot\int d^2\Vang~
\up{Re}\left[\GXY(\Vang)\WY^*(\Vang)\right]
~e^{-i\vL\cdot\Vang}.
\label{eq:GW}
\end{eqnarray}
The normalization coefficients, $A^{XY}_l$, are related to the noise power spectrum, $\Nkap_l$, of the estimators $\khat^{XY}(\Vang)$ by $\Nkap_l=l^2A^{XY}_l/4$, and can be expressed as \bear
\label{eq:nps}
{\frac{1}{ A^{XY}_l}}&=&{\frac{1}{
l^2}}\int{\frac{d^2\vL_1}{(2\pi)^2}}{(\vL\cdot\vL_1)~
\frac{C^{XY}_{l_1}f^{XY}_{\vL_1\vL_2}}{\CXobs_{l_1}~\CYobs_{l_2}}} \\
&\times&
\bigg\{\begin{array}{c}\cos2\Delta\varphi\\\sin2\Delta\varphi\end{array}\bigg\}
~e^{-l_1^2\sbeam^2}~e^{-l_2^2\sbeam^2},\nonumber \enar with
$\vL=\vL_1+\vL_2$, $\Delta\varphi=\varphi_{\vL_1}-\varphi_{\vL_2}$,
and $\langle X_{\vL_1}Y_{\vL_2}\rangle=f_{\vL_1\vL_2}^{XY}~\phi_\vL$~,
where~\cite{Hu:2001tn} \bear
\label{eq:fl}
f^{TT}_{\vL_1,\vL_2}&=&(\vL\cdot\vL_1)~\CT_{l_1}+(\vL\cdot\vL_2)~\CT_{l_2},
\\ f^{TE}_{\vL_1,\vL_2}&=&(\vL\cdot\vL_1)~\CC_{l_1}\cos2\Delta\varphi
+(\vL\cdot\vL_2)~\CC_{l_2}, \nonumber \\
f^{TB}_{\vL_1,\vL_2}&=&(\vL\cdot\vL_1)~\CC_{l_1}\sin
2\Delta\varphi,\nonumber \\
f^{EE}_{\vL_1,\vL_2}&=&\left[(\vL\cdot\vL_1)~\CE_{l_1}+(\vL\cdot\vL_2)~\CE_{l_2}
\right]\cos2\Delta\varphi,\nonumber \\
f^{EB}_{\vL_1,\vL_2}&=&(\vL\cdot\vL_1)~\CE_{l_1}\sin2\Delta\varphi.\nonumber
\enar
 On the basis of the relations above, we stress that a careful estimation of the noise contribution to lensing depends on how accurately the observed spectra are 
known, as well as how much the exponential representation of the high $l$ cutoff due to instrumental beam in (\ref{eq:nps}) is indeed faithful.
Our code for estimating the convergence using the quadratic estimator formalism is a direct implementation of the above equations, \eqn{eq:nois}--(\ref{eq:fl}), and was exploited in 
\citep{Fantaye:2012ha}. In that work, the CMB lensing signal was directly simulated on flat sky. In the present one, we need to project a curved sky onto a flat patch, in order 
to proceed with the analysis. We exploit a gnomonic projection scheme validating it in the next Section. 

\begin{table}[htb]
\begin{center}
\begin{tabular}{rcc}
Experiment & FWHM & $\sigma_{pixT}$ ($\mu$K$\cdot$arcmin)\\
\hline
{\sc Planck} & 7.18' & 43.1\\
\hline 
{\sc PRISM} & 3.2' & 2.43\\
\hline
\hline
\end{tabular}
\caption{{\sc Planck} and {\sc PRISM} performance specifications. Beam FWHM is given in arcminutes, and the sensitivity for $T$ per pixel in $\mu$K$\cdot$arcmin. The channels used are 143 GHz for {\sc Planck} and 160 GHz for {\sc PRISM}.
The polarization sensitivity for both $E$ and $B-$modes is $\sqrt{2}\Delta T/T$.}
\label{table_experiments}
\end{center}
\end{table}

In Fig. \ref{figure_noise} we show the forecasted noise spectra for the minimum variance estimator in a {\sc Planck}-like case \cite{planck} and in a  {\sc PRISM}-like case \cite{prism} (for the adopted specifications see Tab. \ref{table_experiments}), computed for the {\sc Planck} 1 year cosmology \cite{planckparameters}.
 
By comparing the amplitude of the noise contribution between the {\sc Planck} and {\sc PRISM} cases, we can conclude that the precision of the {\sc Planck}  experiment, despite being extremely powerful on the already delivered temperature 
spectrum, does still not allow for a detection with high signal to noise ratio at the large scales targeted in this work, both due to the beam amplitude and to the sensitivity, whereas in the case of a future survey with the {\sc PRISM} specifications the quality of the measurement will improve significantly, permitting to obtain a highly precise reconstruction also at very small angles. For this reason, in this work we will adopt the {\sc PRISM} specifications to address the contribution coming from non-linearities.

\section{The recovered lensing signal}
\label{sec:iv}
\begin{figure}[t]
\begin{center}
\begin{overpic}[width=0.48\textwidth]{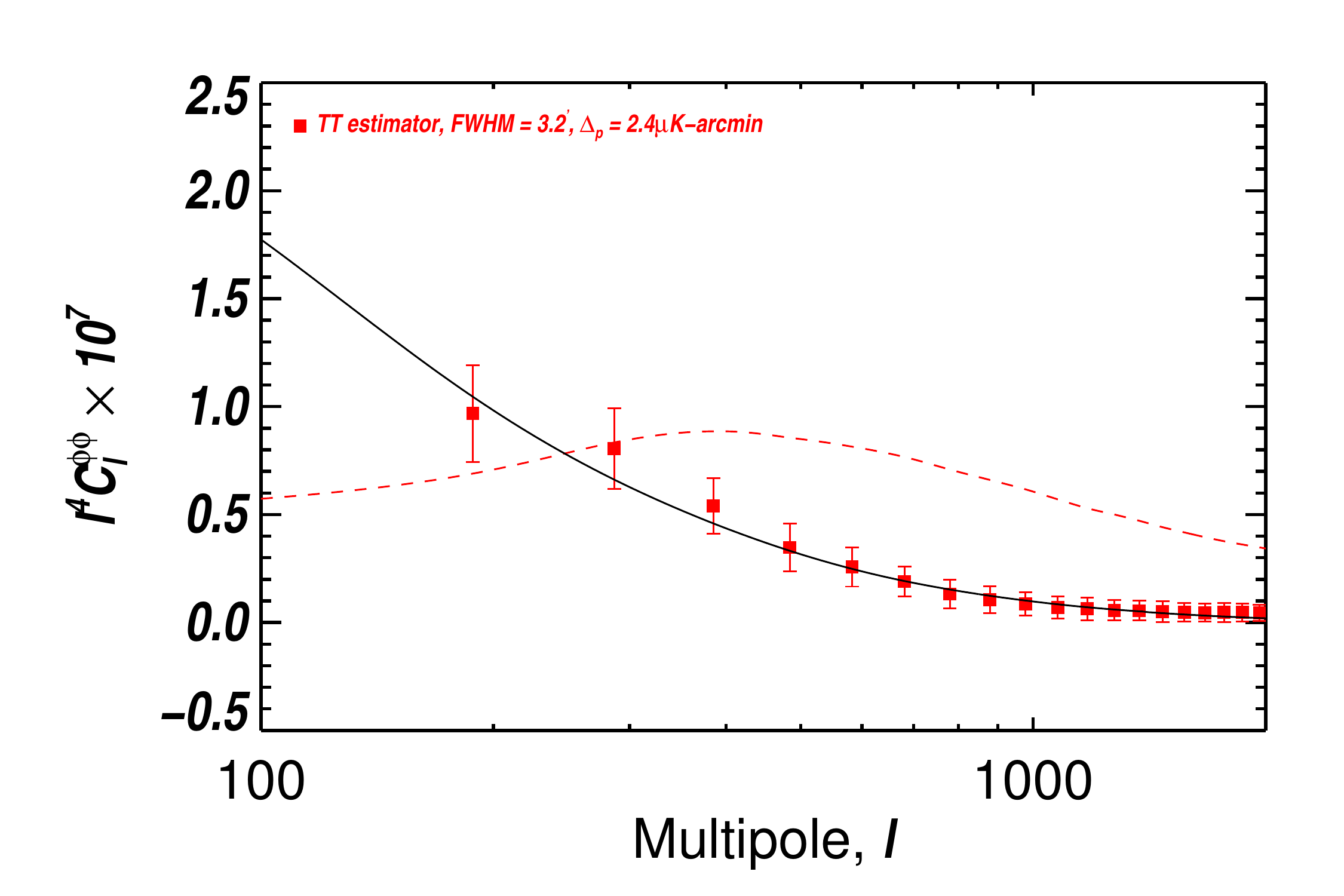}
\put(47,32){\includegraphics[width=.186\textwidth]{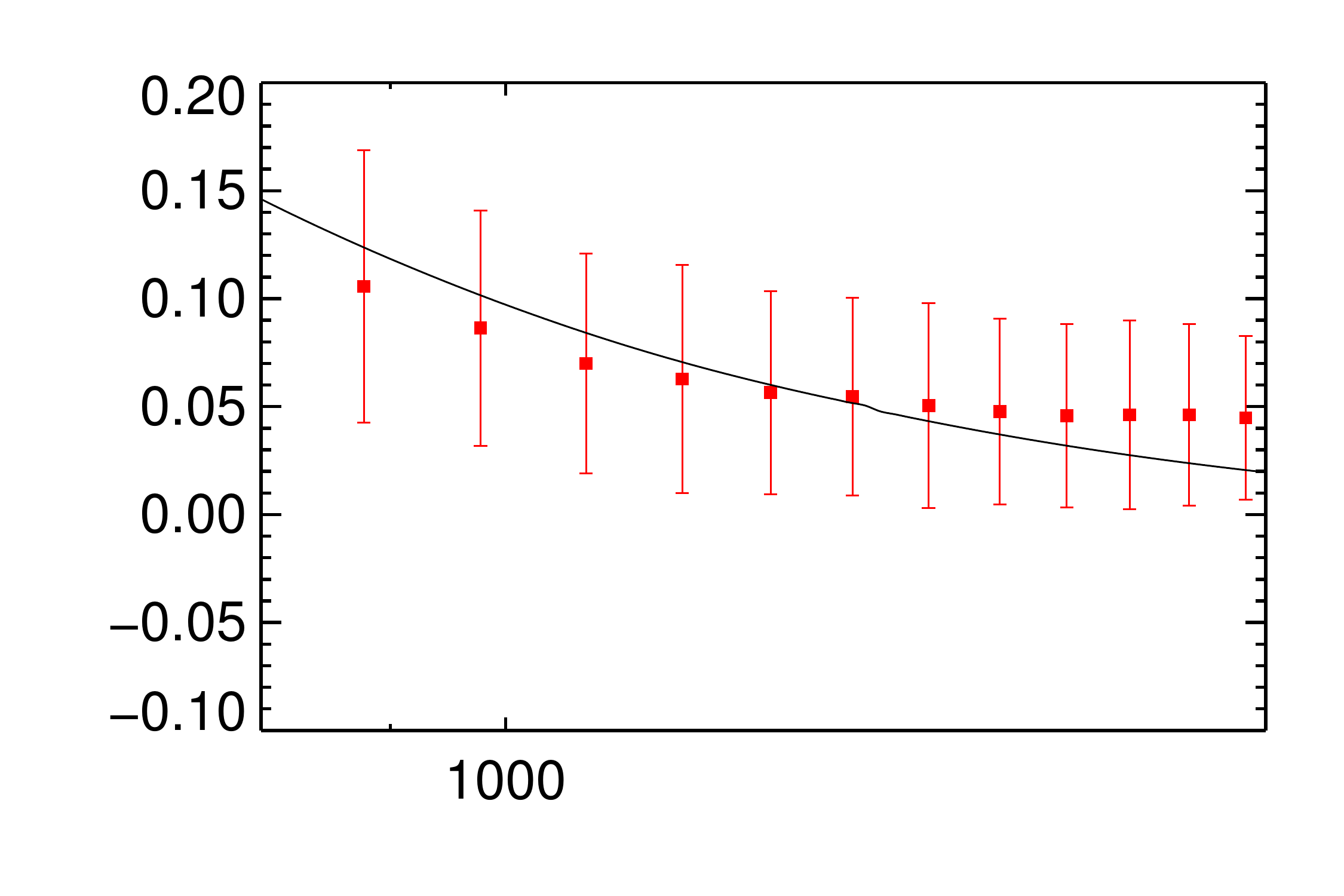}}
\end{overpic}
\begin{overpic}[width=0.48\textwidth]{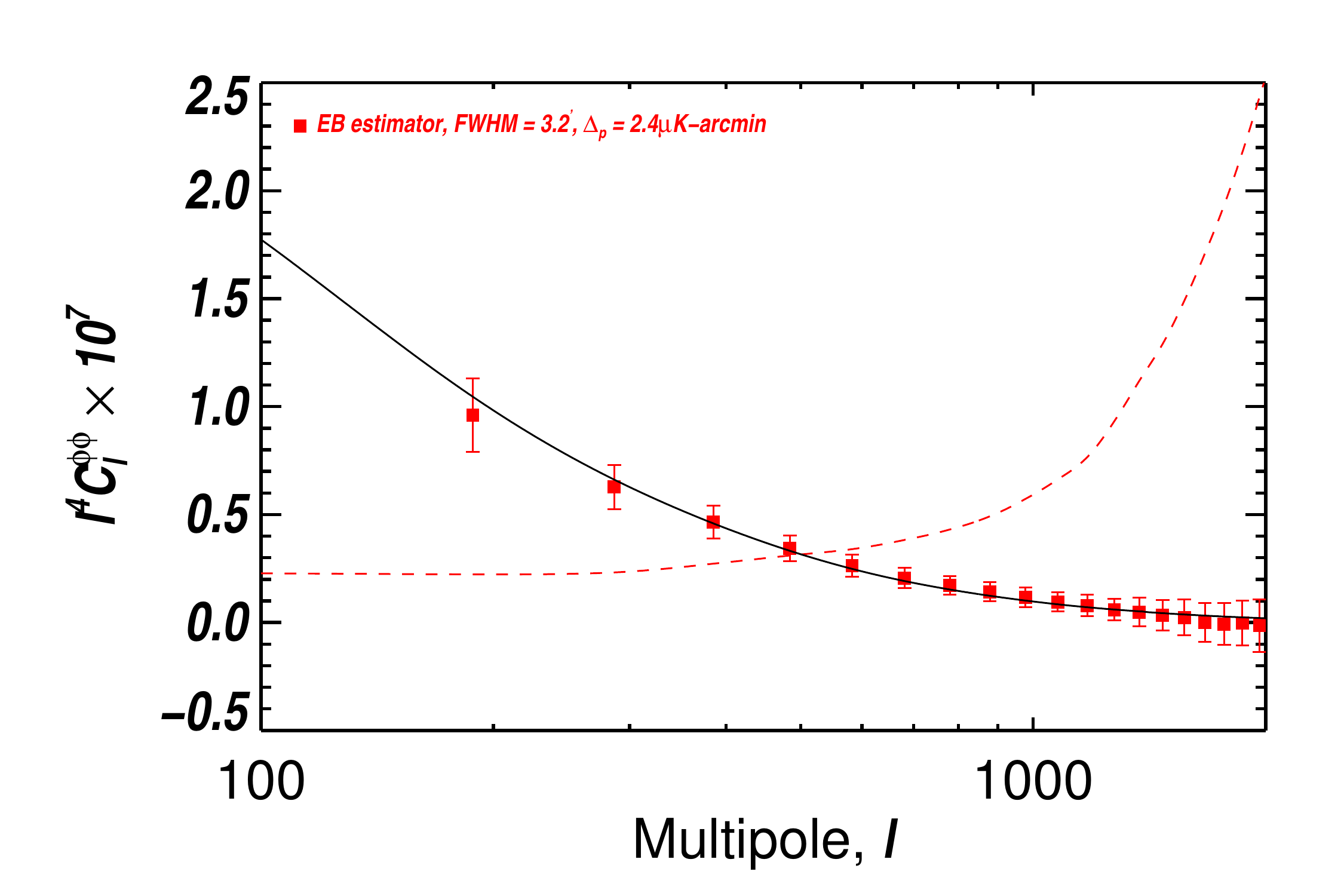}
\put(47,32){\includegraphics[width=.186\textwidth]{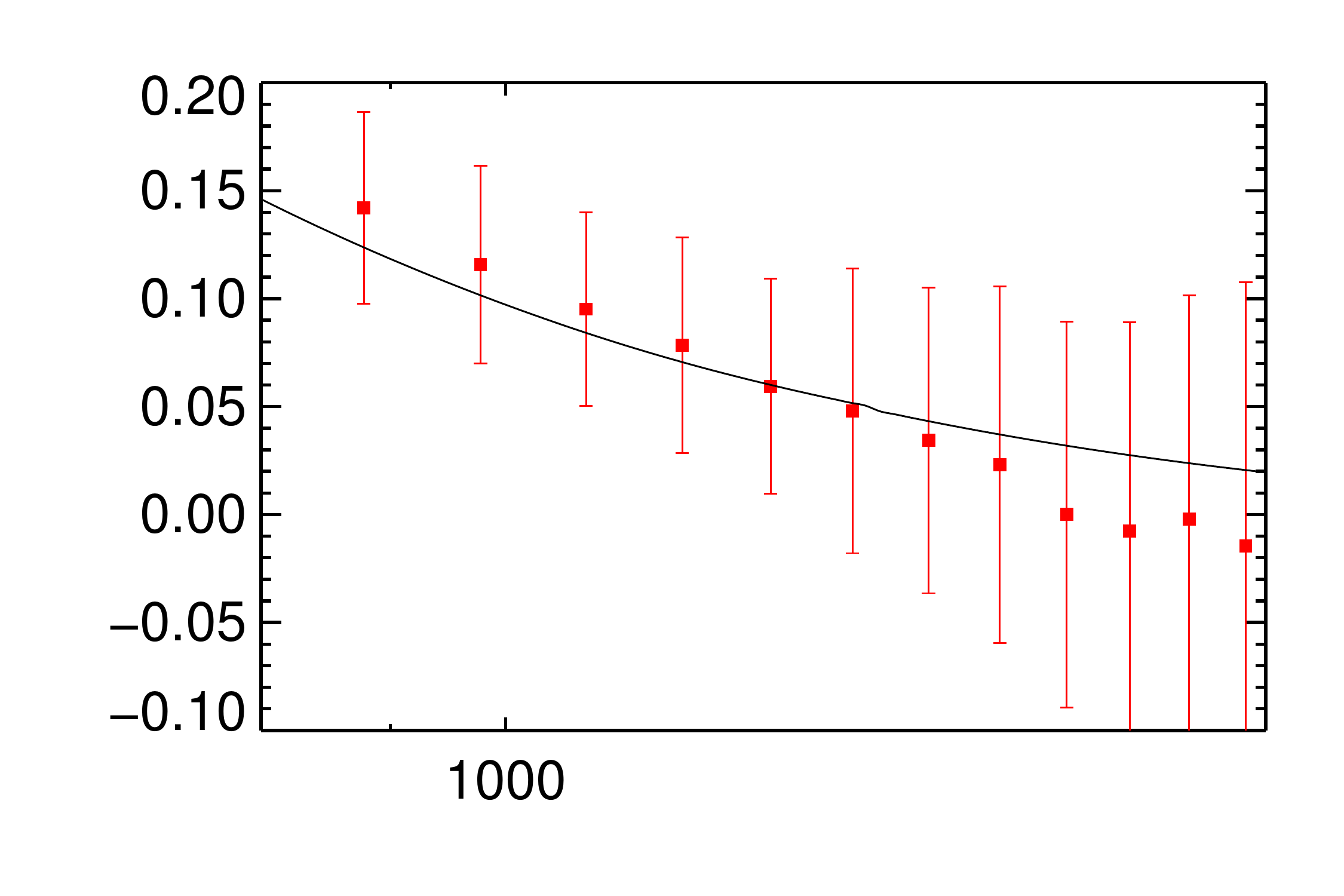}}
\end{overpic}
\caption{Convergence spectrum extraction from LP simulated maps using the $TT$ (left) and $EB$ estimator (right), side of patch side of 15$^{\circ}$ for the {\sc PRISM} 
specifications.  The dashed lines represent the noise contribution as evaluated by the estimator, which has been subtracted from the recovered signal in order to obtain 
the data points. The black line is  the convergence spectrum obtained by {\sc CAMB} for the reference cosmology, the red dashed lines represent the noise  contribution.}
\label{figure_test}
\end{center}
\end{figure}

\begin{figure}[t]
\begin{center}
\begin{overpic}[width=0.48\textwidth]{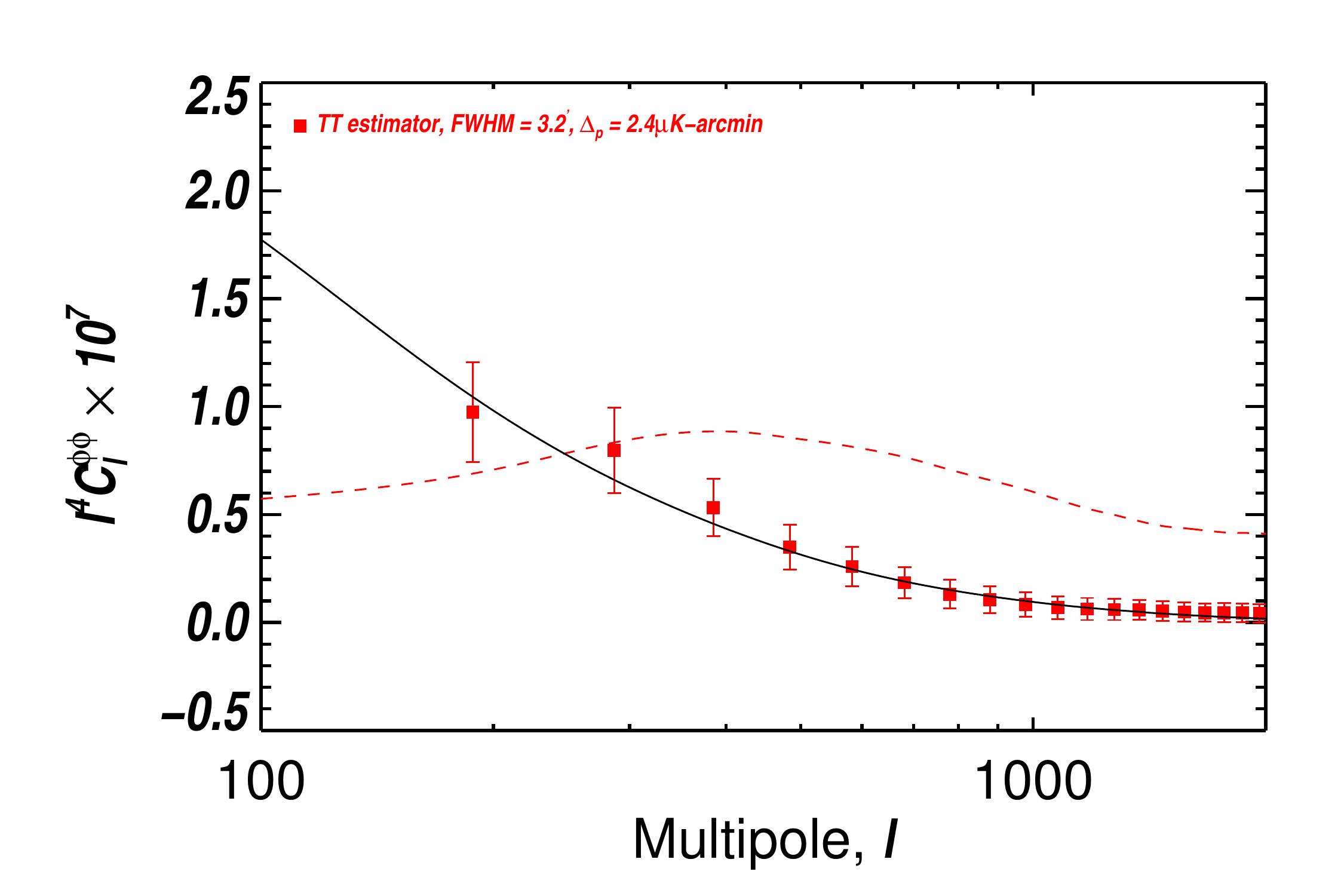}
\put(47,32){\includegraphics[width=.186\textwidth]{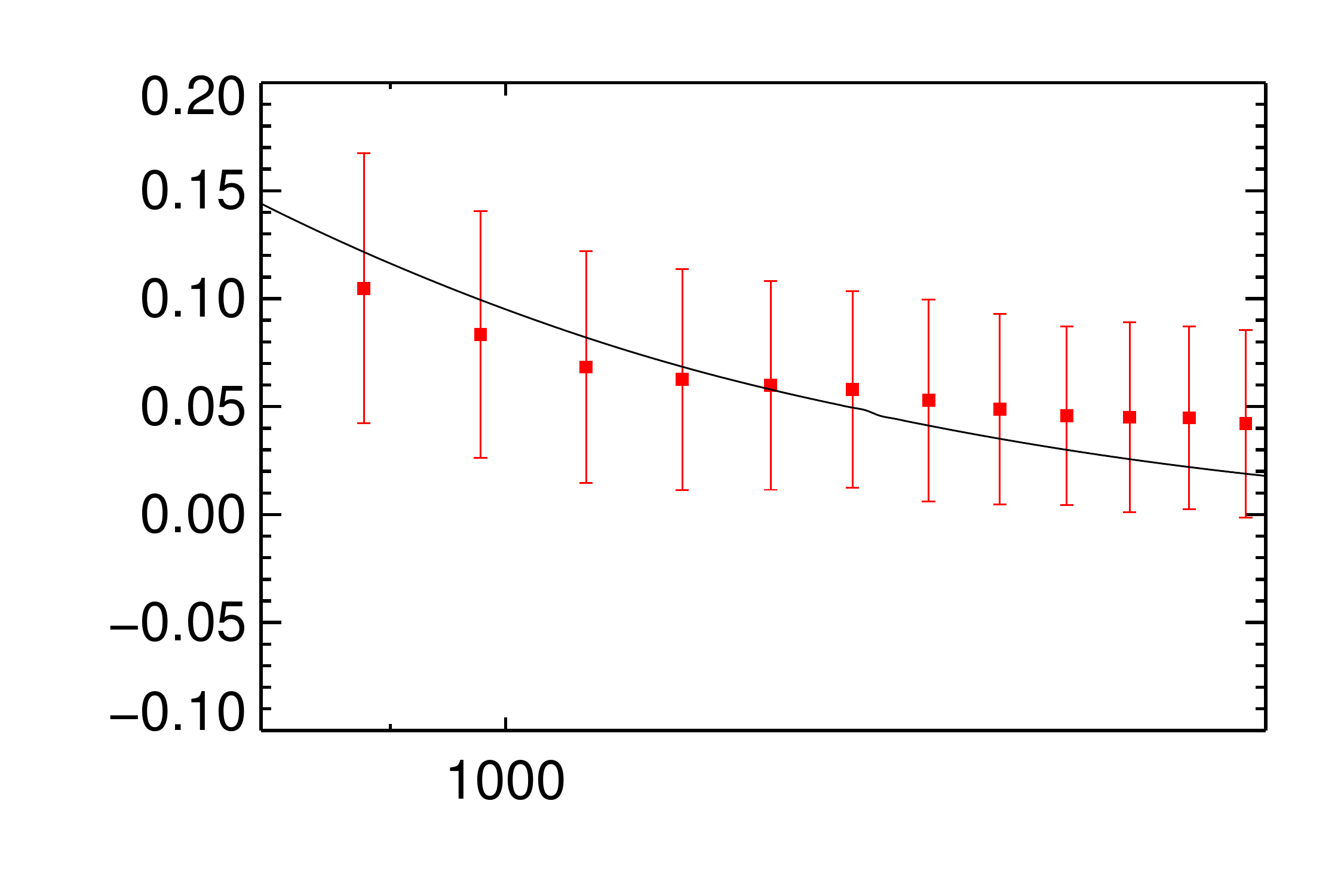}}
\end{overpic}
\begin{overpic}[width=0.48\textwidth]{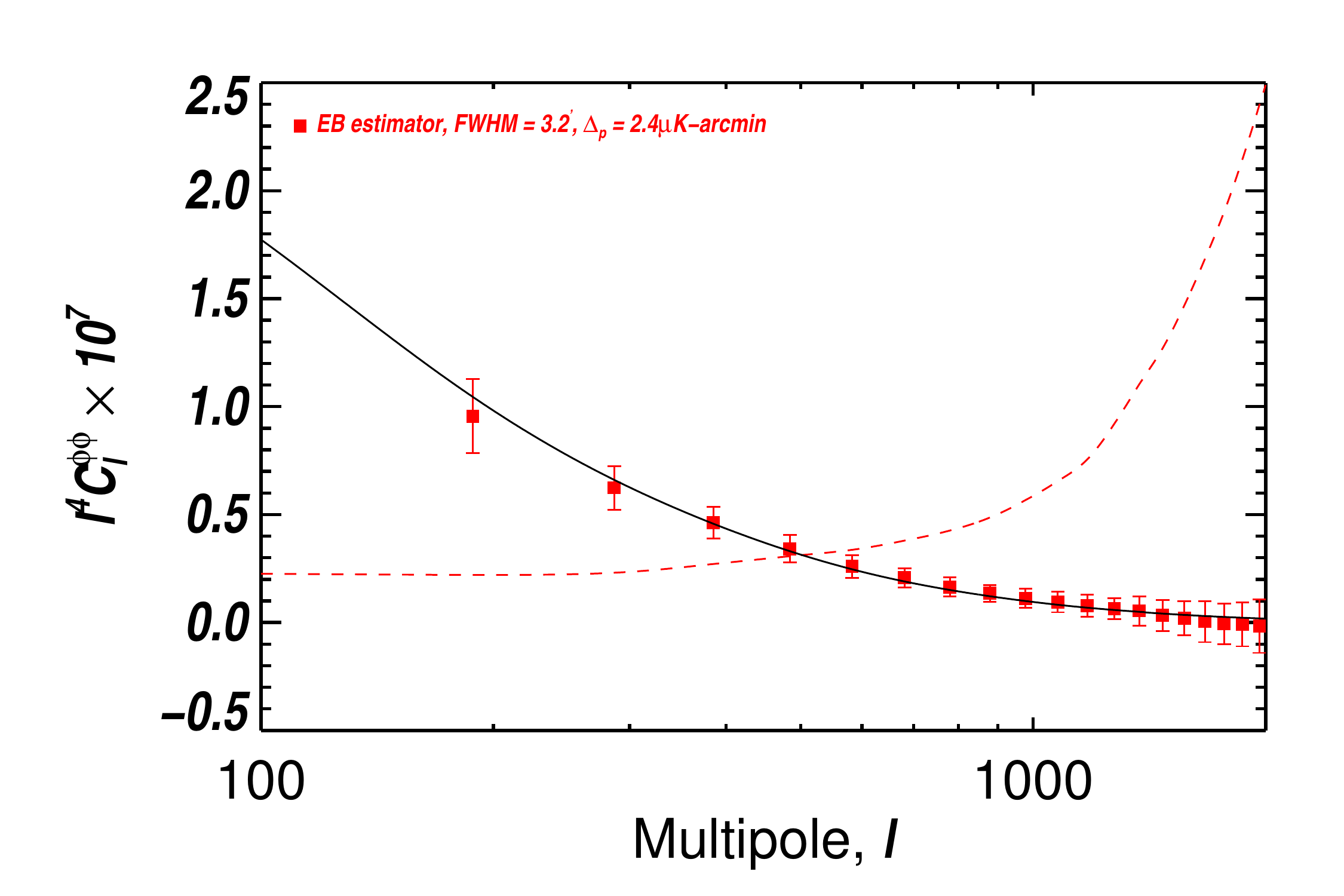}
\put(47,32){\includegraphics[width=.186\textwidth]{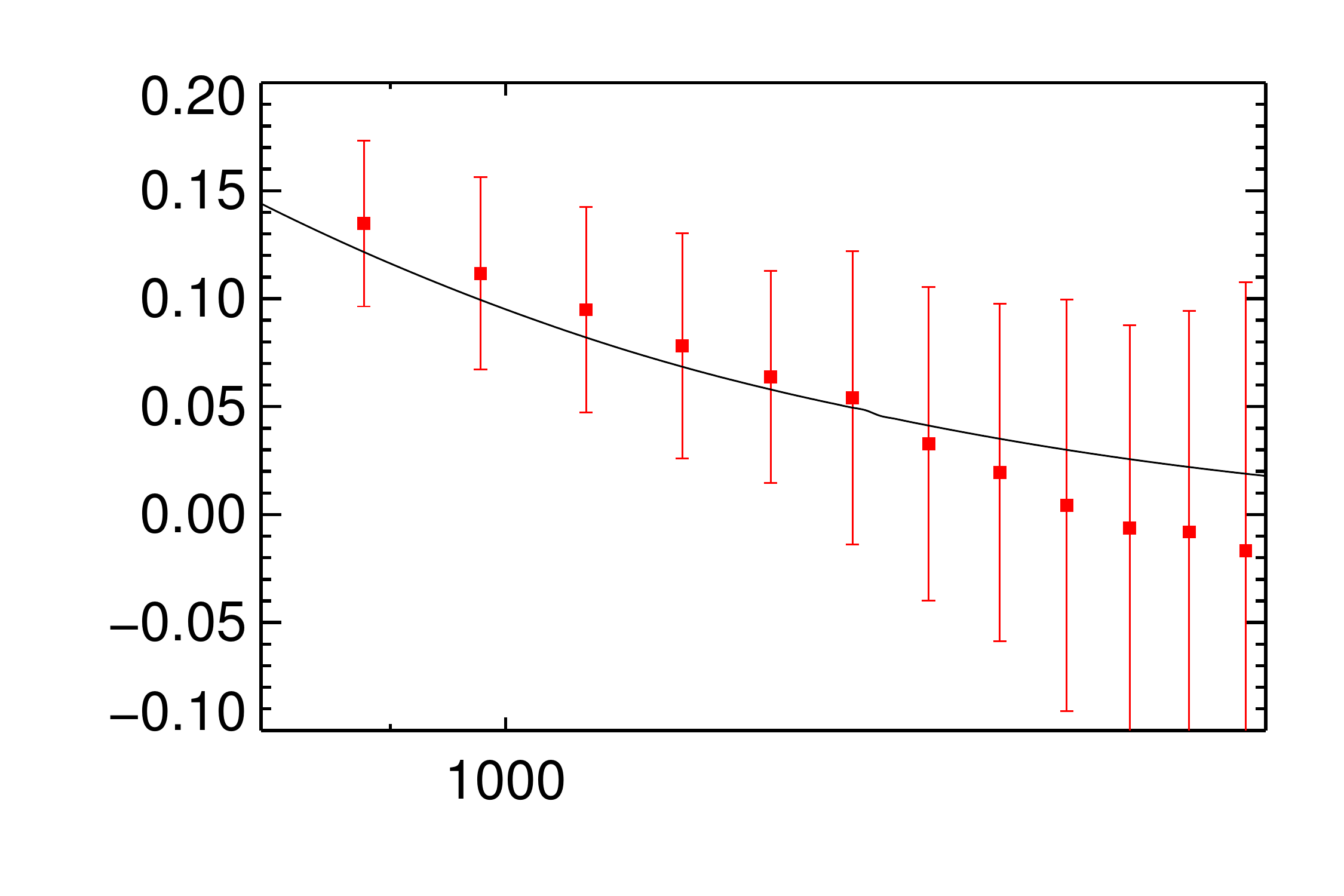}}
\end{overpic}
\caption{Convergence spectrum extraction from N-body lensed maps using the TT (left) and EB estimator (right). Notation and line style associations are the same as in 
Fig. \ref{figure_test}.}
\label{figure_shear_spectra}
\end{center}
\end{figure}

In this Section we discuss the results of our lensing extraction, showing maps of recovered shear lensing signal, and quantitative comparisons of its power spectrum against 
the $\Lambda$CDM predictions in the interval of angular scales which is made accessible by the present simulation setup. First, let us do a few considerations on the noise spectra 
in the angular region of interest. It is known that the noise spectra of all the possible combinations $TT$, $TE$, $TB$, $EE$, $EB$ for the convergence spectrum are relatively flat 
on large scales, just having a difference in amplitude, but not exhibiting a particular dependence  on $\ell $ (see \cite{okamotohu2003}).  As already explained in the previous 
Sections, we are interested in lensing reconstruction ranging from the arcminute to the degree scale, where the noise spectrum is comparable or lower than the signal to extract only 
for the $TT$ and $EB$ cases. Thus, we focus our analysis on these two observables as they are most significant for the experimental configurations we will examine.

We apply the flat sky lensing estimator procedure described in the previous Section by adopting a 15$^{\circ}$ patch side. The lensing extraction pipeline proceeds as follows. From the 
all sky lensed maps, we extract 296 squared patches, with centres distributed following \citep{tiling}. The shear angular power spectra from each single patch 
are then stacked for producing the final result. The noise contribution as predicted by the lensing estimator is subtracted. 
In order to validate our simulation setup, we perform a test run using a simulated LP map by adopting the {\sc PRISM} specifications
and  a \WMAP \ 1 year fiducial set of cosmological parameters \citep{spergel2003}. The resulting convergence spectra  as output by the lensing extraction pipeline and obtained 
by subtracting the noise contribution are shown in Fig. \ref{figure_test} and exhibit 
a complete agreement with the theoretical prediction both for the $TT$ and the $EB$ case. The zoomed regions in the $800\le l\le 2000$ show numerical instabilities which 
are showing up at the highest multipoles. The figure also anticipates some of the features which will be highlighted for the cases of 
the run on the N-body CMB lensed maps, precisely in the shape and amplitude of the noise contributions, for the $TT$ and $EB$ cases. 
The $TT$ case appears to be noise dominated on all angular scales, while the effects of the limited angular resolution are visible at the largest scales in the $EB$ signal, 
in the shape of the noise  contribution, reflected by the error bar increase in the recovered signal at $\ell\gtrsim 1500$. The $\Lambda$CDM predicted power is recovered 
very accurately on all scales, reflecting the precision in the evaluation of the noise  contribution. Finally, with the adopted specifications, the polarised data do represent a significant contribution to the recovery of the signal, with comparable precision up to $\ell\simeq 1500$. It is also interesting to look at the reconstruction precision, reaching a few percent in bins with $\Delta\ell\simeq 100$ in the angular multiple interval $1000\lesssim\ell\lesssim 2000$ for $T$, and about $10\%$ for $EB$.

We now turn to the study of results on the N-body lensed CMB maps. 
The angular power spectra from the shear maps stacking are shown in Fig. \ref{figure_shear_spectra}, where the two panels corresponding 
to the result of the $TT$ (left) and $EB$ (right) estimators, respectively, show the reconstructed lensing potential evaluated by stacking the lensing spectra extracted in each of 
the regions considered.  As expected, the noise contributions for the two cases are the same as the LP case in Fig. \ref{figure_test}. The solid lines corresponding to the 
the spectra after subtraction of the noise contribution show no visible departure from predictions of the weak lensing power as predicted by the $\Lambda$CDM cosmology, within uncertainties, for both cases, in particular on the angular scales which are less affected by Cosmic Variance, e.g. corresponding to $\ell\simeq 1000$ and beyond. 
 The consistency between the two cases keeps validity up to the extreme angular resolution, as it is clear by comparing the zoomed areas in this and \ref{figure_test} cases, 
indicating that the behaviour at the largest scales is actually a numerical feature to be attributed to the estimator rather than to the simulated CMB lensing maps.
It is to be noted that the results in Fig. \ref{figure_shear_spectra} are obtained from the MS-modified-LP as defined in Sec. \ref{sec:ii}, whereas the panels shown in \ref{figure_test} have been obtained with the standard unmodified LP version. This result, validating the whole scheme of 
the simulation pipeline, constructed using N-body structures out of theoretical power predicted semi-analytically, ray traced and then inspected at the level of the CMB lensing 
extraction precision, has the immediate consequence that the biases from inaccuracies across the pipeline are well below the high precision performance of next generation CMB experiments 
for lensing extraction. The outline procedure should then allow to characterize departures from $\Lambda$CDM predictions within the redshift interval which is contributing 
significantly to the lensing power, within the assumed instrumental accuracy. It should be noted that this is true in particular in the small scale part, where the corrections 
from mildly non-linear matter evolution, described through the {\sc Halofit} package into {\sc CAMB} contribute and are faithfully reconstructed. \\

\begin{figure}[t]
\begin{center}
\includegraphics[ trim=7.5cm 2cm 6cm 5cm, clip=true, width=0.49\textwidth]{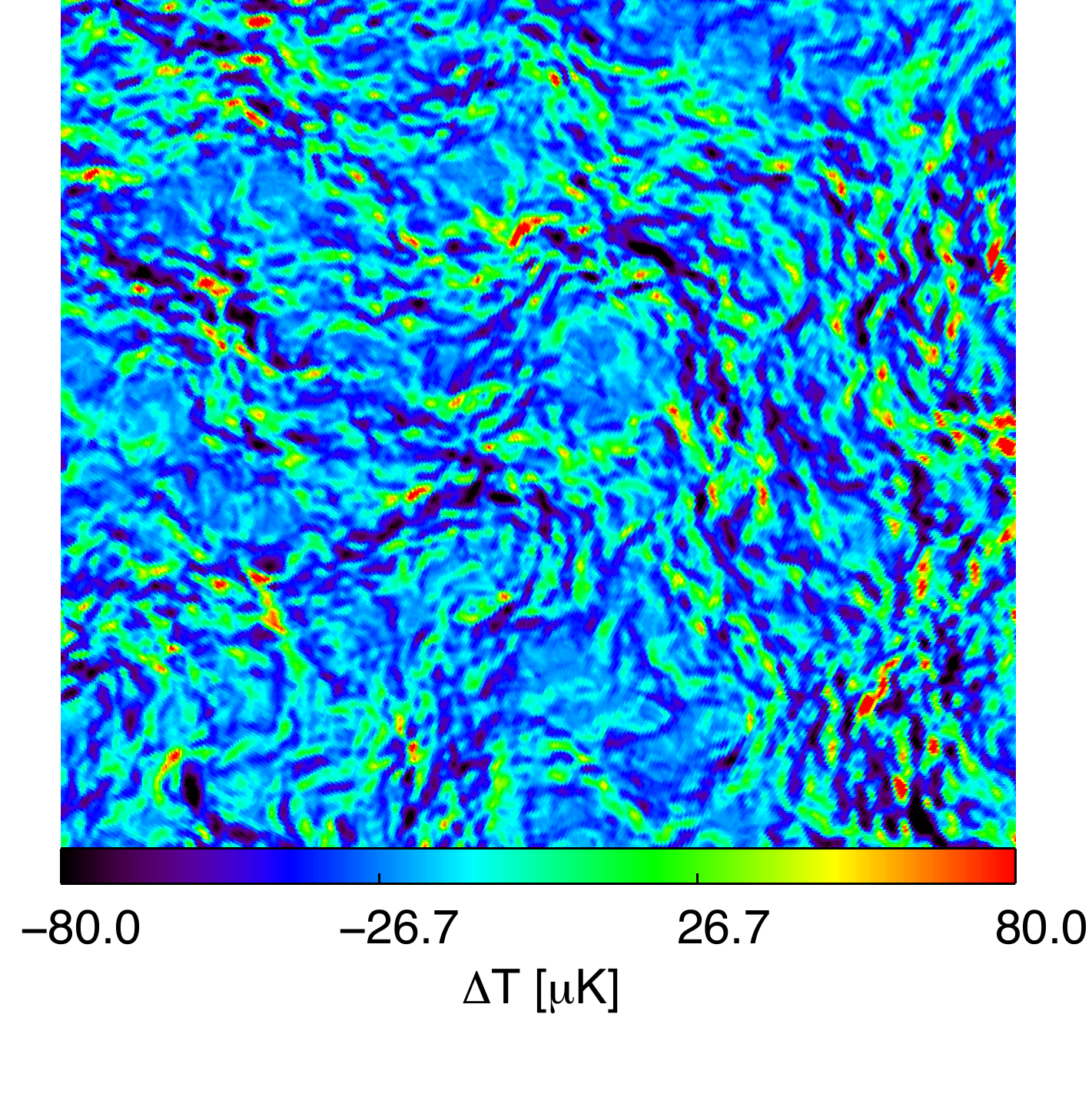}
\includegraphics[ trim=7.5cm 2cm 6cm 5cm, clip=true, width=0.49\textwidth]{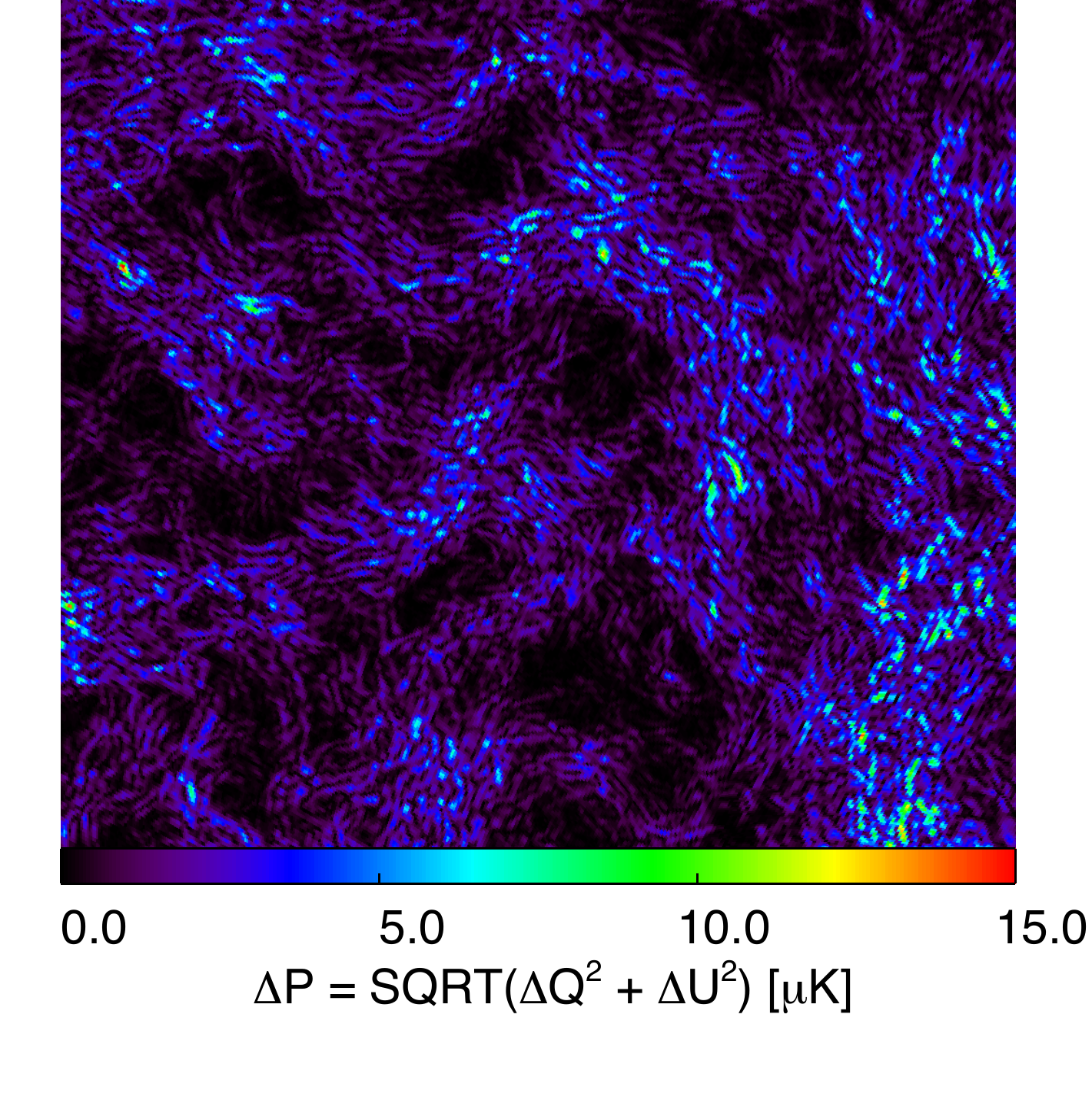}
\includegraphics[trim=7.5cm 2cm 6cm 5cm, clip=true, width=0.49\textwidth]{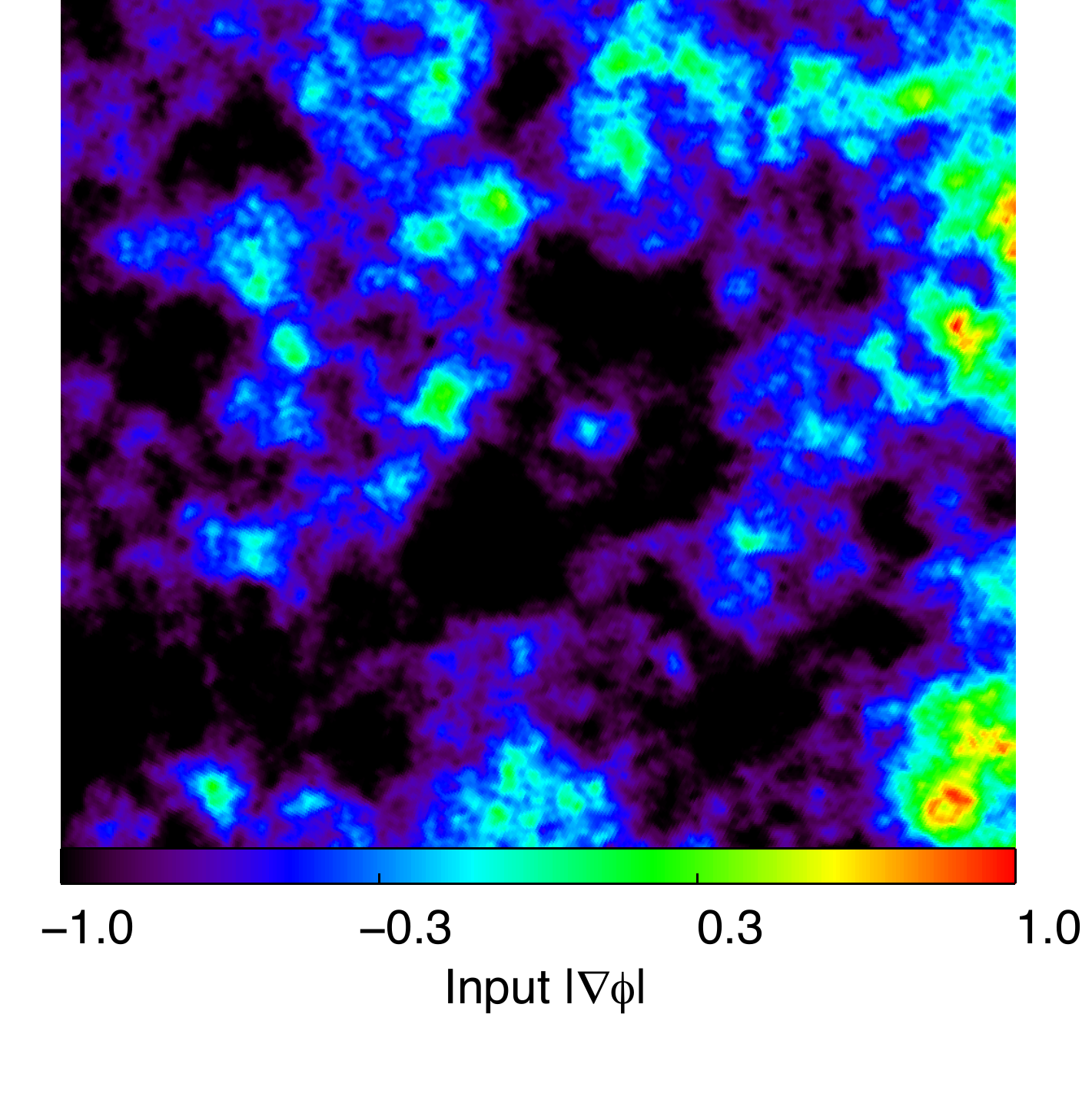}
\includegraphics[trim=7.5cm 2cm 6cm 5cm, clip=true, width=0.49\textwidth]{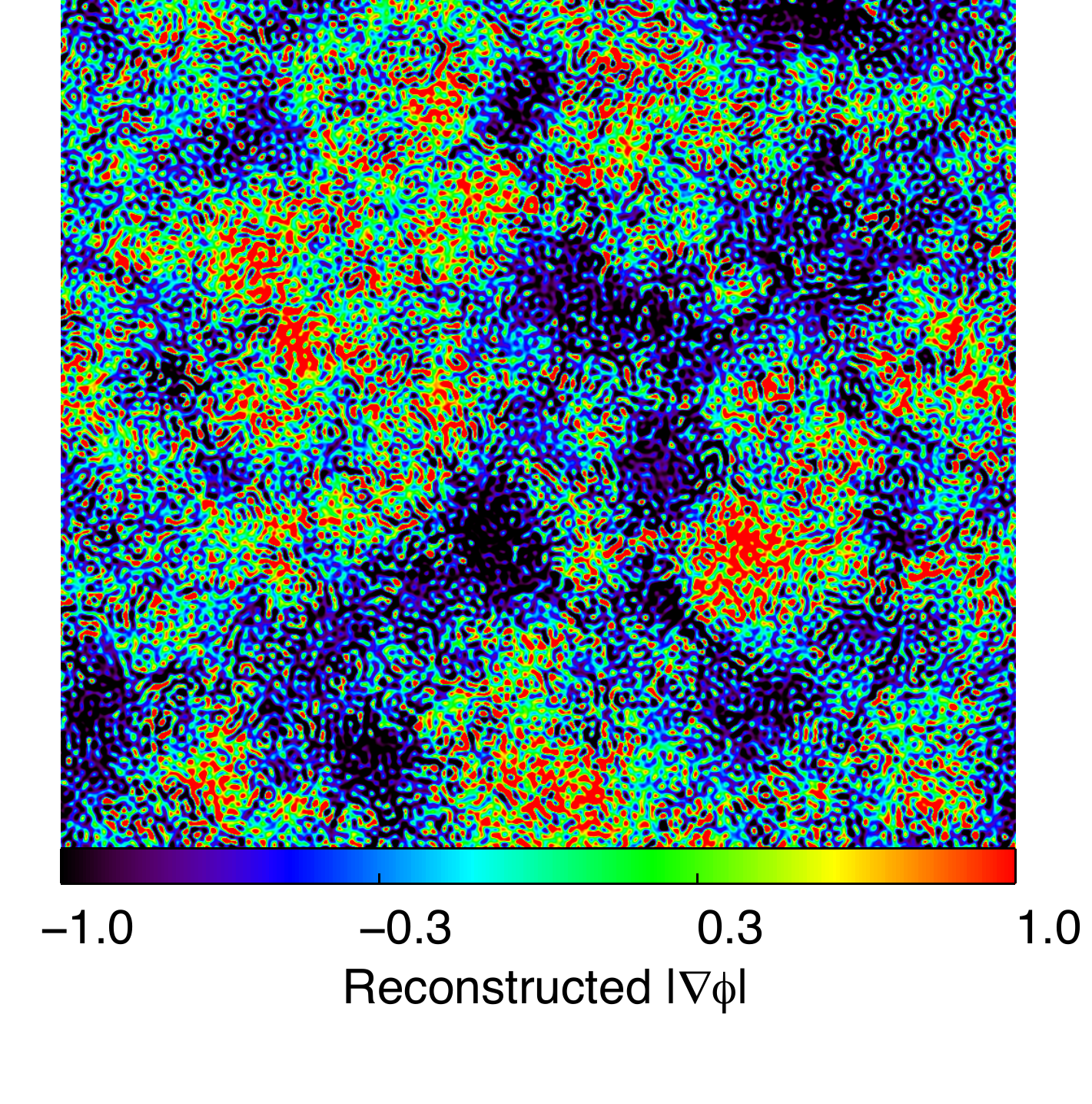}
\caption{Lensed and unlensed difference maps in the temperature (top right panel) and  
polarised intensity fields (top left panel) in the same patch of the CMB maps compared with the modulus of the input (bottom right panel) and reconstructed (bottom left panel) deflection angle.}
\label{figure_shear_T_P_maps}
\end{center}
\end{figure}
Before concluding we perform a last visual study of our results, showing how the lensing signal is consistent in different renderings. In Fig. \ref{figure_shear_T_P_maps} the 
four panels show the modulus of the input and reconstructed deflection angle compared with the difference between the lensed and 
unlensed CMB maps for $T$ and the polarisation amplitude $P=\sqrt{Q^{2}+U^{2}}$. A first immediate evidence is the marked non-Gaussianity of the lensing field, e.g. in 
the $T$ difference; the structures there represent the line of sight integral of MS DM lenses acting on the background CMB field. The same holds for the polarisation field 
difference, with a clear correlation with the $T$ field, as expected, as well as a finer structure in the lensing contribution. The bottom panels show the input and 
reconstructed noisy lensing potential field, again featuring an evident correlation between input and output, depite of the noise pattern, which is also well evident. 
A similar analysis, on the whole sky, was performed by \citep{carbone_etal_2008}, without applying a full lensing extraction pipeline as we do in the present work. 

\section{Concluding remarks}
\label{sec:v}

We presented here the first extraction of lensing shear and quantitative comparison with semi-analytical expectations of Cosmic Microwave Background (CMB) lensing simulations 
obtained through ray-tracing across N-body structure formation. We consider the lensed total intensity and polarisation CMB maps obtained by displacing the background field with 
Born approximated deflection 
angles evaluated from the Millennium Run simulation, stacked to fill up the whole Hubble volume. We test our pipeline by making use of simulated realisations of CMB lensing fields 
where the polarisation angle is assumed to have a Gaussian statistics and a power spectrum as given by semi-analytic predictions. We adopt the specification of future high resolution 
and sensitivity CMB satellites, corresponding to arcminute and $\mu$K-arcminute angular resolution and sensitivity, respectively. The geometry of the simulation setup, corresponding 
to a N-body box size of 500$h^{-1}$ Mpc and a pixelisation with $1.7'$ pixel size, gives us access to angular scales covering the arcminute and reaching about a degree in 
the sky. For that we use a flat sky approximated version of the lensing extraction pipeline based on a quadratic estimator. \\
We inspect separately the extracted lensing pattern from total intensity and polarisation. We discuss the lensing contribution as predicted by the lensing estimator in the two cases, 
finding the signal to be completely signal dominated for total intensity, while the effect of limited angular resolution is evident in the polarisation noise  contribution
at the small scale edge of the relevant interval. 

By applying the extraction pipeline, we find that the reconstructed weak lensing shear power spectra are featureless as in the case of the simulated maps, following the theoretically 
predicted power within the assumed uncertainties, separately for the total intensity and polarisation based estimator. Within the assumed instrumental specifications, we find that 
the polarisation field has comparable relevance in constraining the lensing signal. 

The performed analysis is relevant in the context of the current and planned CMB and LSS large observational campaigns. In this context, galaxy-galaxy and CMB 
lensing are predicted to be most important observables for constraining the dark cosmological components, and the control and reliability of the corresponding simulated signal possesses 
a crucial importance in the forecasting phase, as well as for the interpretation of the data. For this reason, it is important in particular for CMB lensing to gather the different 
pieces of the simulations in a single pipeline and to study the results. This work represents a first significant step in this direction, demonstrating not only that the inaccuracies 
of the simulated cosmological structure, ray tracing scheme and lensing extraction provide no significant disturbance to the lensing recovery on the entire interval of angular scale 
considered, but also that this procedure can be upgraded by adopting more sophisticated simulations, both in terms of general architecture of the N-body and/or ray tracing procedure, as well as underlying cosmologies. These aspects are indeed the subject of our future works in this direction.

\vspace{4cm}

\section*{Acknowledgements}
This work was supported by the INFN PD51 initiative. CB, MM and CA acknowledge support by the Italian Space Agency through the ASI contracts Euclid-IC (I/031/10/0). YF is supported by ERC Grant 277742 Pascal and acknowledges funding from the research council of Norway. CC acknowledges the INAF Fellowships Programme 2010.
Part of this work was conducted at the Institute of Theoretical Astrophysics, University of Oslo. YF and CA acknowledge support from this institution. Also, CA thanks the Institute of Cosmology and Gravitation and the Perimeter Institute for Theoretical Physics for hospitality during this work.

\end{document}